\documentclass[12pt,aps,a4paper,preprint, showpacs, showkeys, superscriptaddress]{revtex4} 
\usepackage{graphicx}
\usepackage[breaklinks=true,colorlinks=true,linkcolor=red,urlcolor=blue,citecolor=green]{hyperref}
\usepackage{amsmath}
\begin{document}

\title[TAAP]{ Different atom trapping geometries with time averaged adiabatic potentials}

\author{Sourabh Sarkar}
\email{sourabhs@rrcat.gov.in}
\affiliation{Laser Physics Applications Section, Raja Ramanna Centre for Advanced Technology, Indore, 452013, India}
\affiliation{Homi Bhabha National Institute, Training School Complex, Anushakti Nagar, Mumbai, 400094, India}
\author{S. P. Ram}
\affiliation{Laser Physics Applications Section, Raja Ramanna Centre for Advanced Technology, Indore, 452013, India}

\author{V. B. Tiwari}
\affiliation{Laser Physics Applications Section, Raja Ramanna Centre for Advanced Technology, Indore, 452013, India}
\affiliation{Homi Bhabha National Institute, Training School Complex, Anushakti Nagar, Mumbai, 400094, India}

\author{S. R. Mishra}
\affiliation{Laser Physics Applications Section, Raja Ramanna Centre for Advanced Technology, Indore, 452013, India}
\affiliation{Homi Bhabha National Institute, Training School Complex, Anushakti Nagar, Mumbai, 400094, India}

\begin{abstract}
In this article, we have theoretically studied the time averaged adiabatic potential (TAAP) scheme for realizing different atom trapping geometries. It is shown that by varying time orbiting potential (TOP) fields and radio frequency (rf) fields parameters, controlled manipulation of trapping potentials, and conversion from one trapping geometry to another, is possible. The proposed trapping geometries can be useful for studying various atom-optic phenomena such as Bose-Einstein condensation (BEC) in low dimensions, super-fluidity, tunnelling, atom interferometry, etc. 

\end{abstract}


\vspace{2pc}

\maketitle

\section{Introduction}
\label{sec:intro}

With the advent of laser cooling and trapping of atoms \cite{Metcalf1999}, the research in atomic physics has evolved in several dimensions from basic science \cite{Albiez2005,Ramanathan2011,Schumm2005} to practical applications \cite{BertoldiA.2006,Peters2001,MuellerT.2009} of cold atoms in quantum technologies \cite{Mark2011,CRYU2013,Alzar2019}. A magneto-optical trap (MOT) \cite{WohllebenW.2001} is a work-horse device to produce samples of cold atoms for various studies and applications. The static magnetic traps \cite{Merloti2013,Chakraborty2016} and optical dipole traps \cite{GRIMM200095} are widely used traps for further trapping and cooling of atoms from MOT to generate ultra-cold or degenerate atomic gas samples. In spite of a standard practice of using these magnetic and dipole traps for cold atoms, it is felt that a good handle on control and manipulation of trapping geometries can enrich our capabilities for use of cold atoms for several purposes. In this context, the rf-dressed adiabatic potential scheme, first proposed by Zobay and Garraway \cite{Zobay2001}, has been implemented successfully to demonstrate atom trapping in different geometries \cite{Chakraborty2014, Heathcote2008,Zobay2001,Hofferberth2006,Sherlock2011}. The rf-dressed magnetic traps \cite{Chakraborty2014, Chakraborty2016, Merloti2013,Easwaran2010} are able to generate some novel trapping geometries such as double well \cite{Schumm2005}, ring trap \cite{Morizot2006, Heathcote2008,Sherlock2011}, arc trap \cite{Chakraborty2014}, etc, which are difficult to achieve by conventional static magnetic traps or laser dipole traps. These capabilities can be further enriched by use of Time Averaged Adiabatic Potential (TAAP) technique, which is an extension of the rf-dressed adiabatic potential technique and proposed by Lesanovsky and Klitzing \cite{Lesanovsky2007}. In the TAAP scheme, a relatively lower frequency time averaging field, called time orbiting potential (TOP) \cite{Petrich1995} field, is applied on atoms in presence of rf and static magnetic fields used for rf-dressing. Gildemeister et al \cite{Gildemeister2012} have first experimentally demonstrated the working of atom trapping using TAAP scheme. 
 
 In this work, we have theoretically investigated various atom trapping geometries using TAAP scheme. Our results show that by varying the TOP fields parameters as well as rf-dressing fields parameters, various atom trapping geometries can be realized which are not possible with rf-dressing alone. In TAAP scheme, the conversion from one geometry to the other is possible. Some of these trapping geometries and their conversions can be used to study the dynamics of cold atoms and Bose-Einstein condensates \cite{Anderson1995,Davis1995} in low dimensions \cite{Merloti2013}. Also, these trapping geometries can find applications in study of tunnelling  \cite{Albiez2005,Schumm2005,Hofferberth2007}, super-fluidity \cite{Ramanathan2011,Heathcote2008,Gildemeister2012}, atom interferometry \cite{Kasevich1991,Peters2001,Canuel2006,Carnal1991}, etc, which have applications in various atom-optic devices. 
 
 The article is organized as following. The section \ref{sec:theory} presents the general theoretical frame work for description of rf-dressed adiabatic potentials and Time Averaged Adiabatic Potential (TAAP). In section \ref{sec:discussion}, we discuss our proposals for atom trapping geometries achievable using TAAP scheme along with the possibility of conversion from one geometry to the other. Finally, in section \ref{sec:conclusion}, we present the conclusion of our studies.
 
\section{Theory}
\label{sec:theory}
When atoms are trapped in the static magnetic trap, the application of an rf-field results in the formation of new states, know as rf-dressed states. The energy eigen values corresponding to these dressed states are known as rf-dressed potentials (or adiabatic potentials). The rf-dressed potentials generally vary with position of the atom in the trap because of position dependent rf-field coupling strength inside the trap. Thus, atoms trapped in  a magnetic trap in a particular magnetic hyperfine state can be transferred to a rf-dressed potential by exposing them to a rf-field of suitable frequency and amplitude \cite{Chakraborty2016}. In the present theoretical work, we consider the quadrupole trap as a static magnetic trap with field variation given as
\vspace{0.2cm}
\begin{equation} \label{stat mag}
\textbf{B}_\textbf{S}(r) = B^{\prime}_{q}\begin{pmatrix}
                                                                       x\\
                                                                       y\\
                                                                       -2z\\
                                                                       \end{pmatrix} ,
\end{equation}
where $B^{\prime}_q$ is the quadrupole field gradient.

We take the rf-field of the form given by

\begin{equation} \label{Ori Mag}
\textbf{B}_{\textbf{rf}}(t)  =    \begin{pmatrix} 
                                                      B^{x}_{rf} \cos(\omega_{rf}t)\\
                                                      B^{y}_{rf} \cos(\omega_{rf}t + \phi^{y}_{rf})\\
                                                      B^{z}_{rf} \cos(\omega_{rf}t + \phi^{z}_{rf})\\
                                                     \end{pmatrix} , 
\end{equation} 
where $B^{x}_{rf}$, $B^{y}_{rf}$ and $B^{z}_{rf}$ are the amplitudes of the rf-field along $x$-, $y$- and $z$- directions respectively,  and $\omega_{rf}$ is the angular frequency of the rf field. The parameters $\phi^{y}_{rf}$ and  $\phi^{z}_{rf}$ are the phases of the $y$- and $z$-components of rf-field with respect to its $x$-component.

For an atom, in a state with hyperfine angular momentum \textbf{F} and interacting with the  fields given by Eq. (\ref{stat mag}) and Eq. (\ref{Ori Mag}), the effective Hamiltonian in absence of any kinetic energy term can be written as
\begin{equation}
H_{eff}(r,t)= - \boldsymbol{\mu}.{\textbf{B}_\textbf{eff}(\textbf{r},t)} = {\frac{g_{F}\mu_{B}}{\hbar}}{\textbf{F}.{\textbf{B}_\textbf{eff}(\textbf{r}, t)}} ,
\end{equation}
with
\begin{equation} 
\textbf{B}_\textbf{eff}(\textbf{r}, t) = \textbf{B}_\textbf{S}(\textbf{r}) + \textbf{B}_{\textbf{rf}}(t) = \mid\textbf{B}_\textbf{S}(\textbf{r})\mid {\boldsymbol{\hat{e}}}_{S} + \textbf{B}_{\textbf{rf}}(t) ,
\end{equation}
where $\hat{e}_S$ is the direction of the static magnetic field and $\boldsymbol{\mu} = - \frac{g_{F}\mu_{B}}{\hbar}\textbf{F}$ is the magnetic dipole moment of the atom having hyperfine angular momentum \textbf{F}. 

For simplicity of calculations, a new co-ordinate system is chosen whose one of the axes coincides with the direction of quadrupole magnetic field. In this co-ordinate system, the static field vector becomes ${{(0, 0, \mid \boldsymbol{B_{S}(r)} \mid)}^{T}}$, and the rf-field vector becomes ${(B_{rf}^{\perp 1}cos(\omega_{rf}t), B^{\perp 2}_{rf}cos(\omega_{rf}t + \gamma_{1}), B^{\parallel}_{rf}cos(\omega_{rf}t + \gamma_{2}))}^{T}$, where $\gamma_{1}$ and $\gamma_{2}$ are the relative phases of the field components in the new co-ordinate system. The hyperfine angular momentum is also changed to ${(F^{\perp 1}, F^{\perp 2}, F^{\parallel})}^{T}$. Now, if the Schr{\"o}dinger wave equation is transformed in a rotating frame using an unitary operator described as $U = exp[i\omega_{rf}t(\frac{F^{\parallel}}{\hbar})]$, then the effective Hamiltonian in the rotating frame can be written as

\begin{equation} \label{Effective Hamiltonian}
H_{eff}^{R}(t) = UH_{eff}{U}^{\dagger} - i\hbar U \frac{\partial{U}^{\dagger}}{\partial t} =\frac{g_{F}{\mu}_{B}}{\hbar}\Big[F^{\parallel}B_{S} + F^{\parallel}B_{rf}^{\parallel} + \textbf{F}_\textbf{R}.\boldsymbol{{B}_{rf}^{\perp}} -\frac{\hbar \omega_{rf}}{g_{F}\mu_{B}}F^{\parallel} \Big] ,
\end{equation}
where, $F^{\parallel}$ is the angular momentum along the static field direction. The parameter $\textbf{F}_\textbf{R}$ is the angular momentum perpendicular to the static field direction in the rotating frame. The components of $\textbf{F}_\textbf{R}$ can be written as ${F}_{R}^{\perp 1} = U F^{\perp 1} U^{\dagger}$ and ${F}_{R}^{\perp 2} = U F^{\perp 2} U^{\dagger}$.
\vspace{0.2cm}

The first term in Eq. (\ref{Effective Hamiltonian}) is the Zeeman energy term and second term involves the component of rf field along the static field. Second term has a vanishing contribution because $\mid\mu_{B}B_{rf}^{\parallel}\mid\ll\hbar{\omega}_{rf}$. After inserting the expression \cite{Gildemeister2010a} for $\textbf{F}_{\textbf{R}}$ and $\boldsymbol{{B}_{rf}^{\perp}}(t)$, and applying the rotating wave approximation, the expression for the effective Hamiltonian can be written as
\begin{equation}
H_{eff}^{R}=\frac{g_{F}{\mu}_{B}}{\hbar}\Big[F^{\parallel}\Big(B_{S} - \frac{\hbar \omega_{rf}}{g_{F}{\mu}_{B}}\Big) + \frac{F_{+}}{4}\Big(B_{rf}^{\perp 1} - iB_{rf}^{\perp 2}e^{i{\gamma}}\Big) + \frac{F_{-}}{4}\Big(B_{rf}^{\perp 1} + iB_{rf}^{\perp 2}e^{-i{\gamma}}\Big)\Big] ,
\end{equation}
where $F_{+} = F^{\perp 1} + iF^{\perp 2}$ and $F_{-} = F^{\perp 1} - iF^{\perp 2}$ are the raising and lowering operator of hyperfine angular momentum in the rotated frame. $\gamma = \gamma_{2} - \gamma_{1}$ is the relative phase separation of the rf-field components in rotated co-ordinate system.\\ 

The eigenvalues of the Hamiltonian ($H_{eff}^{R}$), when solved in the basis of $ |F, m_{F}\rangle$ states, leads to the expression for the  rf-dressed adiabatic potential as
\begin{equation}
E_{AP}(r)=m_{F} \hbar \sqrt{{\Big(\frac{g_{F} \mu_{B} |\boldsymbol{B_{S}(r)}|}{\hbar} - {\omega}_{rf}}\Big)^2 + \Big(\frac{g_{F} \mu_{B}}{2 \hbar}\Big)^2 \Big[{\Big(B_{rf}^{\perp 1}\Big)}^2 + {\Big(B_{rf}^{\perp 2}\Big)}^2 +2B_{rf}^{\perp 1}B_{rf}^{\perp 2} \sin(\gamma)\Big]}.
\end{equation} 

In a compact form, the adiabatic potential can be expressed as
\begin{equation}
E_{AP}(r) = m_{F}\hbar\sqrt{\mid{\Delta(r)\mid}^2 + {|\Omega_{R}(r)|}^2} \label{Adiabatic},
\end{equation}

 where detuning ($\Delta$) and Rabi frequency ($\Omega_{R}$) for the dressing rf-field are expressed as
\begin{equation}
\Delta(r) = \frac{g_{F}{\mu}_{B}}{\hbar}|\boldsymbol{B_{S}(r)}| - \omega_{rf} .  \label{Detuning}
\end{equation}
and\\
\begin{equation}
{|\Omega_{R}(r)|}^2 = {\Big({\frac{g_{F}{\mu}_{B}}{2\hbar}}\Big)}^2\Big[{\Big(B_{rf}^{\perp 1}\Big)}^2 + {\Big(B_{rf}^{\perp 2}\Big)}^2 + 2B_{rf}^{\perp 1}B_{rf}^{\perp 2} \sin(\gamma) \Big] .
\end{equation}

Using the reverse transformation for fields, the expression for the Rabi frequency ($\Omega_{R}(r)$) in terms of field parameters in lab frame is given by
\begin{multline} \label{Rabi Frequency}
{\mid\Omega_{R}(r)\mid}^{2}  = {\Big(\frac{{{g}_{F}{\mu}_{B}}}{2\hbar}\Big)}^{2}\Bigg[\frac{4{z}^2}{x^2 + y^2 + 4{z}^2}\Bigg(\frac{{\Big(B_{rf}^{x}\Big)}^2 {x}^2 + {\Big(B_{rf}^{y}\Big)}^{2} {y}^{2} }{{x}^{2} + {y}^{2}}\Bigg) + \Bigg(\frac{{\Big(B_{rf}^{x}\Big)}^{2} {y}^{2} + {\Big(B_{rf}^{y}\Big)}^{2} {x}^{2}}{{x}^{2} + {y}^{2}}\Bigg) \\
+ {\Big(B_{rf}^{z}\Big)}^{2}\Bigg(\frac{{x}^{2} + {y}^{2}}{{x}^{2} + {y}^{2} + 4{z}^{2}}\Bigg) - \frac{2 {B}_{rf}^{x} {B}_{rf}^{y} x y \cos({\phi}_{rf}^{y})}{{x}^{2} + {y}^{2} +4{z}^{2}} + \frac{4 {B}_{rf}^{x} {B}_{rf}^{y} z \sin({\phi}_{rf}^{y})}{\sqrt{{x}^{2} + {y}^{2} +4{z}^{2}}} + \frac{4 {B}_{rf}^{y} {B}_{rf}^{z} y z \cos({\phi}_{rf}^{y} - {\phi}_{rf}^{z})}{{x}^{2} + {y}^{2} +4{z}^{2}} \\
+ \frac{2 {B}_{rf}^{y} {B}_{rf}^{z} x \sin({\phi}_{rf}^{y} - {\phi}_{rf}^{z})}{\sqrt{{x}^{2} + {y}^{2} +4{z}^{2}}} + \frac{4 {B}_{rf}^{z} {B}_{rf}^{x} x z \cos({\phi}_{rf}^{z})}{{x}^{2} + {y}^{2} +4{z}^{2}} + \frac{2 {B}_{rf}^{z} {B}_{rf}^{x} y \sin({\phi}_{rf}^{z})}{\sqrt{{x}^{2} + {y}^{2} +4{z}^{2}}}\Bigg].
\end{multline}
From above expression (Eq. (\ref{Adiabatic})) of adiabatic potential, it is evident that potential energy for atom trapping can be tailored to a large extent by choosing the phases ($\phi_{rf}^{y},\phi_{rf}^{z}$) and the amplitudes ($B_{rf}^{x}, B_{rf}^{y}, B_{rf}^{z}$) of the rf-fields.

The time averaged adiabatic potentials (TAAP) can be achieved by time averaging the rf-dressed or adiabatic potentials. This is achieved by applying the low frequency time orbiting potential (TOP) field in addition to dressing rf field to perform the time averaging. 

The general expression of a TOP field is given as
\begin{equation}
\boldsymbol{B_{T}}(t) = B_{T}^{x}\sin(\omega_{T}t)\boldsymbol{\hat{e}}_{x} + B_{T}^{y}\sin(\omega_{T}t + \phi_{T}^{y})\boldsymbol{\hat{e}}_{y} + B_{T}^{z}\sin(\omega_{T}t + \phi_{T}^{z})\boldsymbol{\hat{e}}_{z} ,
\end{equation}
where $B_{T}^{x}$, $B_{T}^{y}$ and $B_{T}^{z}$ are the magnitudes of TOP fields along $x$-,$y$- and $z$-directions respectively. The parameters $\phi_{T}^{y}$ and $\phi_{T}^{z}$ are the phases of the $y$- and $z$-components of TOP field with respect to the $x$- component.

In conventionally used TOP traps, a circularly polarized TOP field in the $x$-$y$ plane is used and its frequency ($\omega_{T}$) is very much greater than the frequency ($\omega_{r}$) of center of mass motion of the atom in quadrupole trap but very much lesser than the Larmor frequency ($\frac{g_{F}{\mu}_{B}B_{T}}{\hbar}$) provided by the TOP field. Mathematically the criteria can be written as
\begin{equation}
\omega_{r}\ll\omega_{T}<\frac{g_{F}{\mu}_{B}B_{T}}{\hbar} ,
\end{equation} 
where $B_{T}$ is the magnitude of the TOP field. 

Finally, the most general expression \cite{Gildemeister2010} of time averaged adiabatic potential (TAAP) can be expressed as
\begin{equation}
\label{TAAPEqn}
E_{TAAP}(r) = \frac{\omega_{T}}{2\pi}\int_{0}^\frac{2\pi}{\omega_{T}}E_{AP}\Big(x + \frac{B_{T}^{x}}{{B}_{q}^{\prime}}\sin(\omega_{T}t), y + \frac{B_{T}^{y}}{{B}_{q}^{\prime}}\sin(\omega_{T}t + \phi_{T}^{y}), z + \frac{B_{T}^{z}}{2{B}_{q}^{\prime}}\sin(\omega_{T}t + \phi_{T}^{z})\Big) dt . 
\end{equation} 

Here it is evident that by changing the TOP field parameters ($B_{T}^{x}, B_{T}^{y}, B_{T}^{z}, \phi_{T}^{y}, \phi_{T}^{z}$) and also the rf field parameters ($B^{x}_{rf}, B^{y}_{rf}, B^{z}_{rf},\phi^{y}_{rf}, \phi^{z}_{rf}$), we can have several interesting atom trapping geometries. Some of these geometries are discussed in the following section.

\section{Results and discussion}
\label{sec:discussion}

 Different atom trapping geometries under TAAP scheme can be achieved with choosing the rf-fields and TOP fields parameters appropriately. Using the TAAP scheme, a vertical ($z$-direction) double well trap was first demonstrated by Gildemeister et al \cite{Gildemeister2010} by choosing a vertical direction ($z$-direction) linearly polarized rf-field $\textbf{B}_\textbf{rf}(t) = {(0, 0, {B}_{rf}^{z}cos(\omega_{rf}t))}^{T}$ and a $x$-$y$ circularly polarized TOP field. Similarly, a ring type atom trap using TAAP scheme was first proposed by Lesanovsky et al \cite{Lesanovsky2007} and experimentally demonstrated by Sherlock et al \cite{Sherlock2011} by applying a circularly polarized rf-field in $x$-$y$ plane and $z$-direction linearly polarized TOP field. The double well trap is useful to study quantum tunneling \cite{Albiez2005} whereas a ring trap can be useful to study the persistent flow \cite{Heathcote2008,Ramanathan2011} of cold atomic gas  and quantum degeneracy \cite{Sherlock2011} in low dimensions. A ring trap geometry is also useful for  matter-wave interferometry \cite{Navez_2016} with cold atoms  which finds applications in developing atom-gyroscope \cite{Alzar2019}. 
 
 In this work,  we have theoretically shown that the variation in rf-fields and TOP fields parameters can give rise to interesting TAAP traps for trapping of cold atoms in different geometries. We have considered that $^{87}Rb$ atoms are magnetically trapped in $|F =2, m_{F} = 2\rangle$ hyperfine level of the ground state $5S_{1/2}$. In the calculations, we have chosen the quadrupole magnetic field with gradient of $B_{q}^{\prime}=$  $100$ G/cm and the rf-field of frequency $\omega_{rf} = $ $2\pi \times 1.5$ MHz. We have actually evaluated Eq. (\ref{TAAPEqn}) to know the trap potential in TAAP scheme for different sets of rf and TOP fields parameters. The discussion on the obtained results is as follows. 

\subsection{Linearly polarized rf-field and TOP field modulations}

\begin{figure}[h!]
\centering
\includegraphics[scale = 0.25]{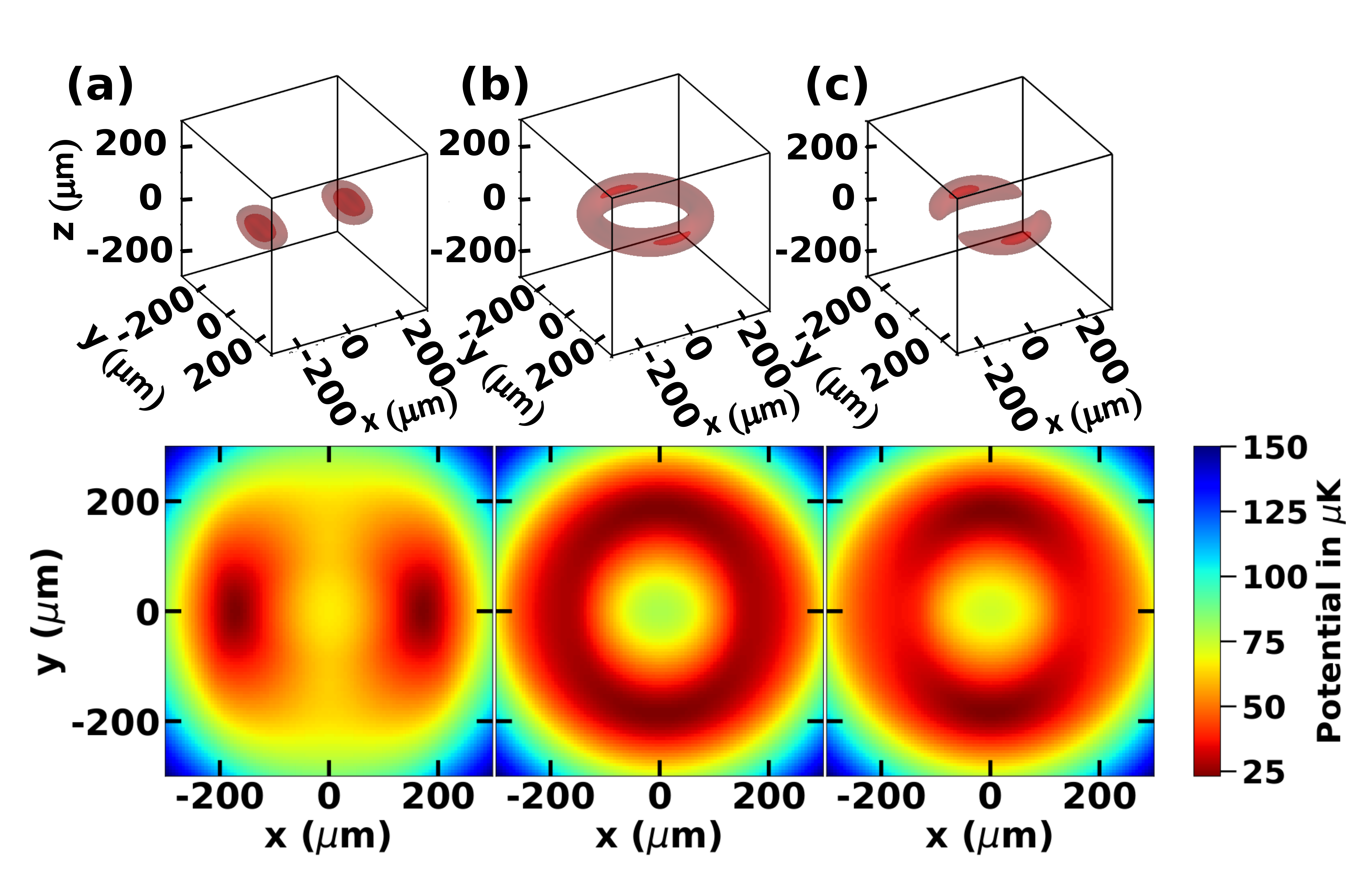}
\caption{(Color online) The conversion of a x-directional double well TAAP trap into a y-directional double well TAAP trap. The upper plots show the 3-D view of $25$ $\mu$K (intense red) and $35$ $\mu$K (light red) iso-potential surfaces, whereas lower plots show the 2-D contours of TAAP potentials for these geometries. The rf-field is linearly polarized in $x$-direction. TOP field is $y$-$z$ circularly polarized ($B_{T}^{y}$ = $B_{T}^{z}$ = $1.3$ G ) for figure (a). For plots (b) and (c) the TOP field is $x$-$z$ polarized with a phase difference of $\frac{\pi}{2}$ between $x$- and $z$- components. The double well gets oriented along $y$-direction for plot (c). The values of TOP fields for plot (b) are $B_{T}^{z}$ = $1.3$ G and $B_{T}^{x} = 650$ mG, and for plot (c) are $B_{T}^{z}$ = $1.3$ G and $B_{T}^{x} = 850$ mG. The other parameters are $B_{rf}^{x}= 700$ mG, $\omega_{rf} = 2\pi \times 1.5$ MHz, $\omega_{T} = 2\pi \times 7$ kHz and ${{B}_{q}^{\prime}} = 100$ G/cm.}
\label{X-y conversion}
\end{figure}

It is known that the double well trap can be generated by using a linearly polarized rf-field and a circularly polarized TOP field in TAAP scheme \cite{Gildemeister2010}. The polarization direction of rf-field has to be perpendicular to TOP field plane for getting the double well oriented along the direction of applied rf-field polarization. With this analogy, a $x$-directional double well can be generated by using a $x$-polarized rf-field and a $y$-$z$ circularly polarized TOP field. Using our simulations, we obtained these results which are shown in Fig. \ref{X-y conversion} (a) via three dimensional (3-D) and two-dimensional (2-D) iso-potential surfaces. We have further investigated that this $x$-directional double well trap can be changed to $y$-directional double well trap by changing only the TOP field configuration to $x$-$z$ polarized. The double well in $y$-direction clearly appears when $B_{T}^{x}$ is increased beyond a certain value for a fixed value of $B_{T}^{z}$. These results are shown in Fig. \ref{X-y conversion} (b) and (c). Here We note that in TAAP scheme, the orientation of double well can be changed by just controlling the TOP field parameters and without any change in rf-field configuration. Double well trap is considered useful for studying tunnelling \cite{Albiez2005} and matter-wave interference \cite{Schumm2005} phenomena. The well separation can be tuned by changing the rf-field frequency $\omega_{rf}$ and quadrupole field gradient ${{B}_{q}^{\prime}}$.

\subsection{Circularly polarized rf-fields and TOP field modulations}

\begin{figure}[h!]
 \centering
 \includegraphics[scale = 0.25]{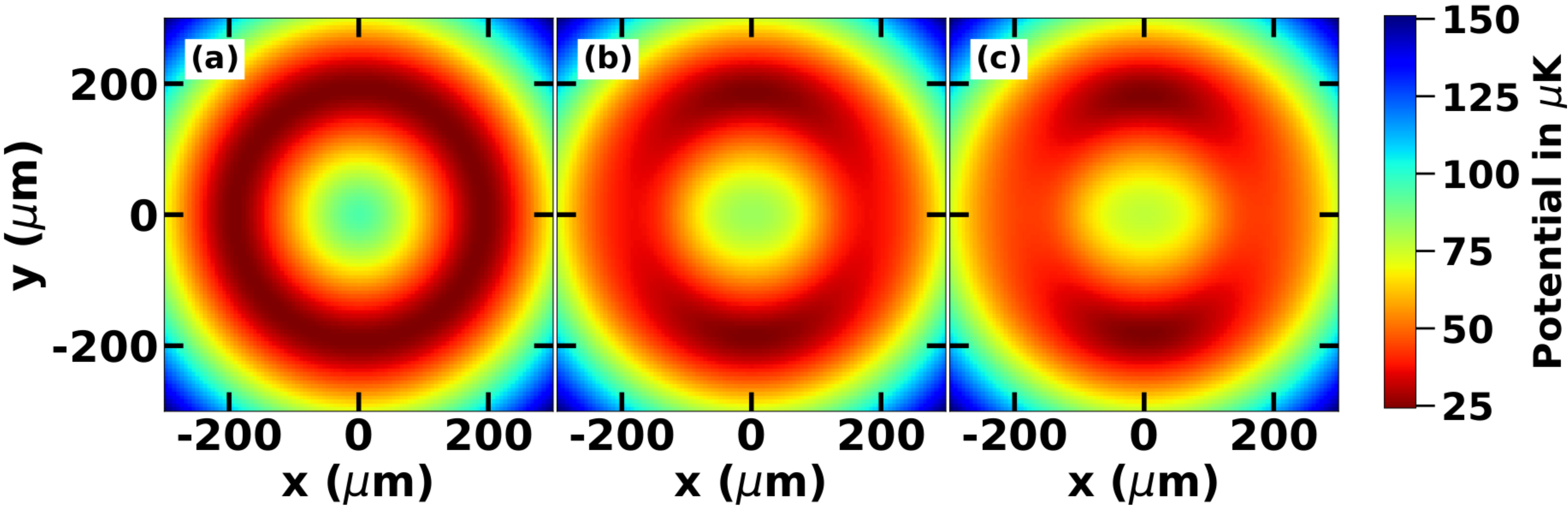}
 \caption{(Color online)  The iso-potential contours of TAAP showing conversion of a ring type trapping potential into a double well along $y$-direction. The circularly polarized rf-field has $B_{rf}^{x} = B_{rf}^{y} = 700$ mG. The value of $z$- TOP field ($B_{T}^{z}$) is $1.3$ G for plot (a). For plot (b), the Values of TOP field components are $B_{T}^{z}$ = $1.3$ G and $B_{T}^{x} = 650$ mG, and for plot (c), $B_{T}^{z}$ = $1.3$ G and $B_{T}^{x} = 850$ mG. The other parameters are $\omega_{rf} = 2\pi \times 1.5$ MHz, $\omega_{T} = 2\pi \times 7$ kHz and ${{B}_{q}^{\prime}} = 100$ G/cm.}
 \label{Ring to ydw}
 \end{figure}

\begin{figure}[h!]
 \centering
 \includegraphics[scale = 0.25]{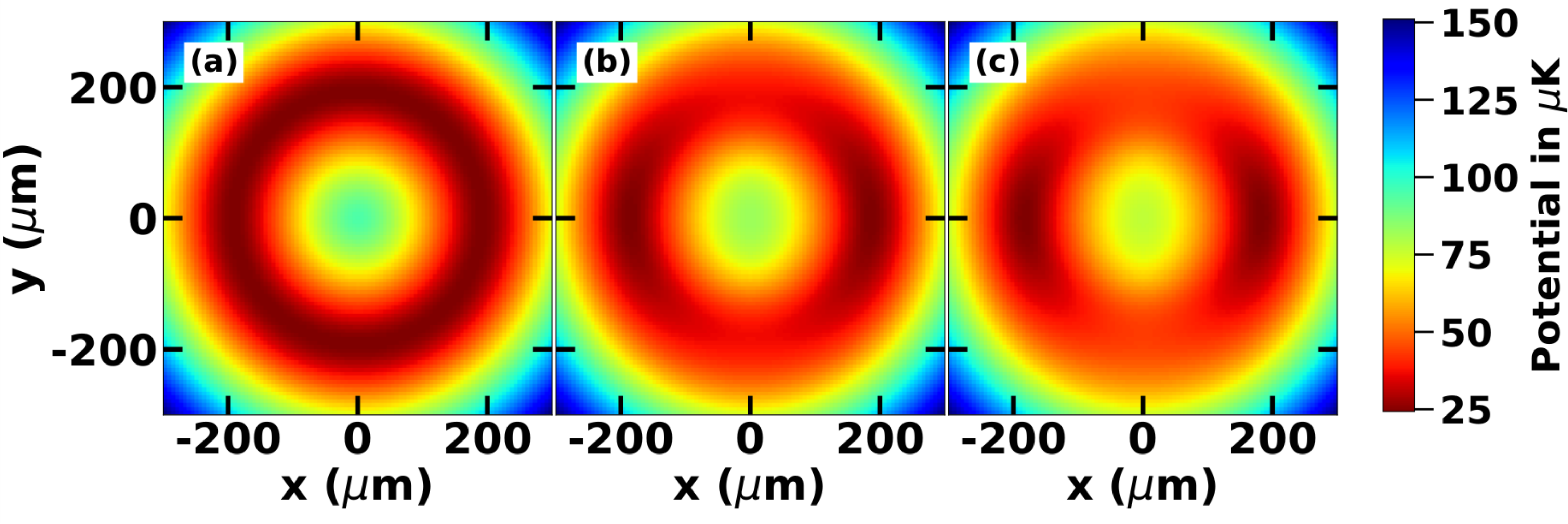}
 \caption{(Color online) The iso-potential contours of TAAP showing conversion of a ring type trapping potential into a double well along $x$-direction. The circularly polarized rf-field has $B_{rf}^{x} = B_{rf}^{y} = 700$ mG. The value of $z$- TOP field ($B_{T}^{z}$) is $1.3$ G for plot (a). For plot (b), the values of TOP field components are $B_{T}^{z}$ = $1.3$ G and $B_{T}^{y} = 650$ mG, and for plot (c), $B_{T}^{z}$ = $1.3$ G and $B_{T}^{y} = 850$ mG. The other parameters are $\omega_{rf} = 2\pi \times 1.5$ MHz, $\omega_{T} = 2\pi \times 7$ kHz and ${{B}_{q}^{\prime}} = 100$ G/cm.}
 \label{Ring to xdw}
 \end{figure}

  It is know from earlier studies \cite{Lesanovsky2007,Sherlock2011},  as well as from simulations here, that a circularly polarized rf-field and a TOP field can give rise to a ring type atom trapping potential in TAAP scheme. For example, a $x$-$y$ circularly polarized rf-field and $z$-polarized TOP field results in a ring TAAP trap in $x$-$y$ plane. With further investigations, we have found that this ring trap can be converted to a $y$-directional double well trap by changing the $z$-polarized TOP field to $x$-$z$ polarized TOP field with a phase difference of $\frac{\pi}{2}$ between $x$- and $z$- components, as shown in  Fig. \ref{Ring to ydw}. The value of phase between $x$- and $z$- components determines the orientation of the double well trap. Similarly, the $x$-$y$ circularly polarized rf-field and $y$-$z$ polarized TOP field can give rise to double well trap along $x$-axis as shown in Fig. \ref{Ring to xdw}. The above kind of conversions from ring to double well may be useful to study the dynamics of super-fluidity \cite{Ramanathan2011} and tunnelling \cite{Albiez2005} with the Bose-condensate of atoms.   
 
\subsection{Multiple rf-fields and TOP field modulations}

\begin{figure}[h!]
 \centering
 \includegraphics[scale = 0.25]{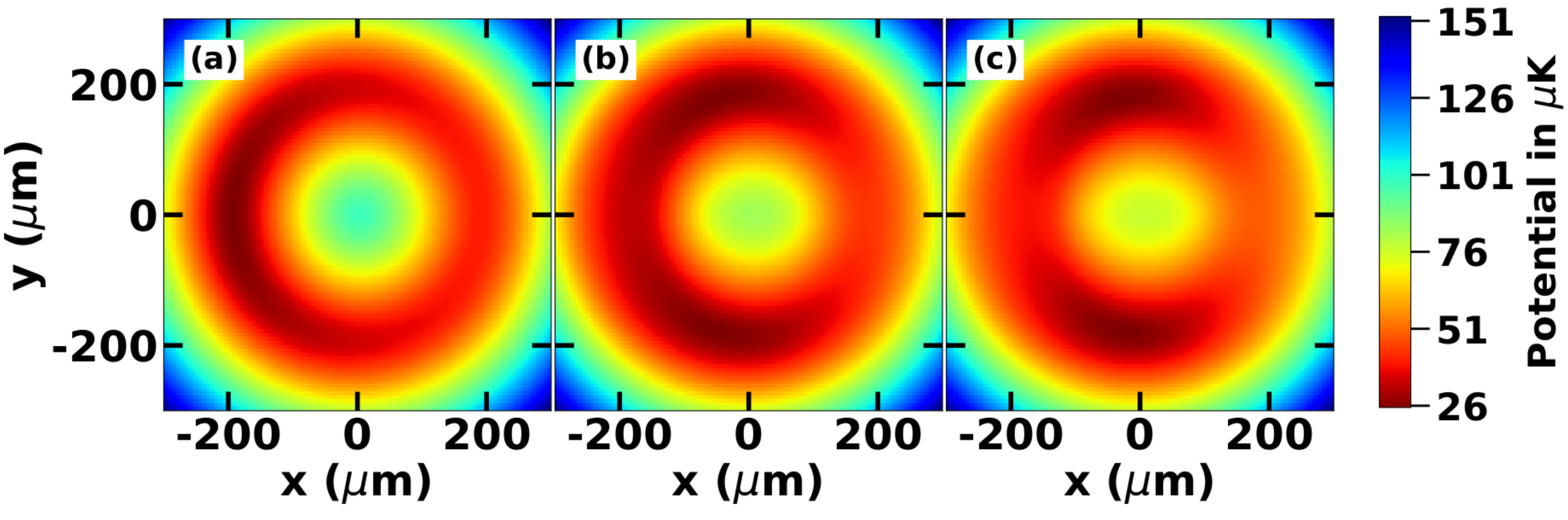}
 \caption{(Color online) The iso-potential contours of TAAP showing an arc type trapping potential and its further conversion into a $y$-directional double well trap. The $x$-$y$ circularly polarized rf-field with $B_{rf}^{x} = B_{rf}^{y} = 700$ mG and a linearly $z$-polarized rf-field with $B_{rf}^{z} = 300$ mG are considered. The value of $B_{T}^{z}$ is taken as $1.3$ G for all plots (a), (b) and (c). Only $z$-TOP field is considered for plot (a), where as $x$-$z$ TOP fields are considered for plots (b) and (c). The values of $B_{T}^{x}$ are $B_{T}^{x} = 650$ mG for plot (b) and $B_{T}^{x} = 850$ mG for plot (c). The other parameters are $\omega_{rf} = 2\pi \times 1.5$ MHz, $\omega_{T} = 2\pi \times 7$ kHz and ${{B}_{q}^{\prime}} = 100$ G/cm.}
\label{Arc to ydw}
\end{figure}

Here we show that use of multiple rf-fields along with TOP fields results in different trapping potentials. For example, a circularly polarized rf-field in $x$-$y$ plane and a $z$-directional rf-field (along with $z$-TOP field) result in a potential minimum on an arc of the ring. The results are shown in Fig. \ref{Arc to ydw}(a). Further more, this arc type trapping potential can be converted into a $y$-directional double well when the $z$-TOP field is changed to a $x$-$z$ TOP field with a phase difference of $\frac{\pi}{2}$ between $x$- and $z$-components. These results are shown in Fig \ref{Arc to ydw}(b) and (c) where $B_{T}^{x}$ is given values of $650$ mG and $850$ mG for (b) and (c) respectively, and $B_{z}^{T} = 1.3$ G is used for both plots (b) and (c). As shown in Fig. \ref{Arc to ydw}(c), the double well potential is better resolved when $B_{T}^{x}$ is given a higher value. 
 
\begin{figure}[h!]
 \centering
 \includegraphics[scale = 0.25]{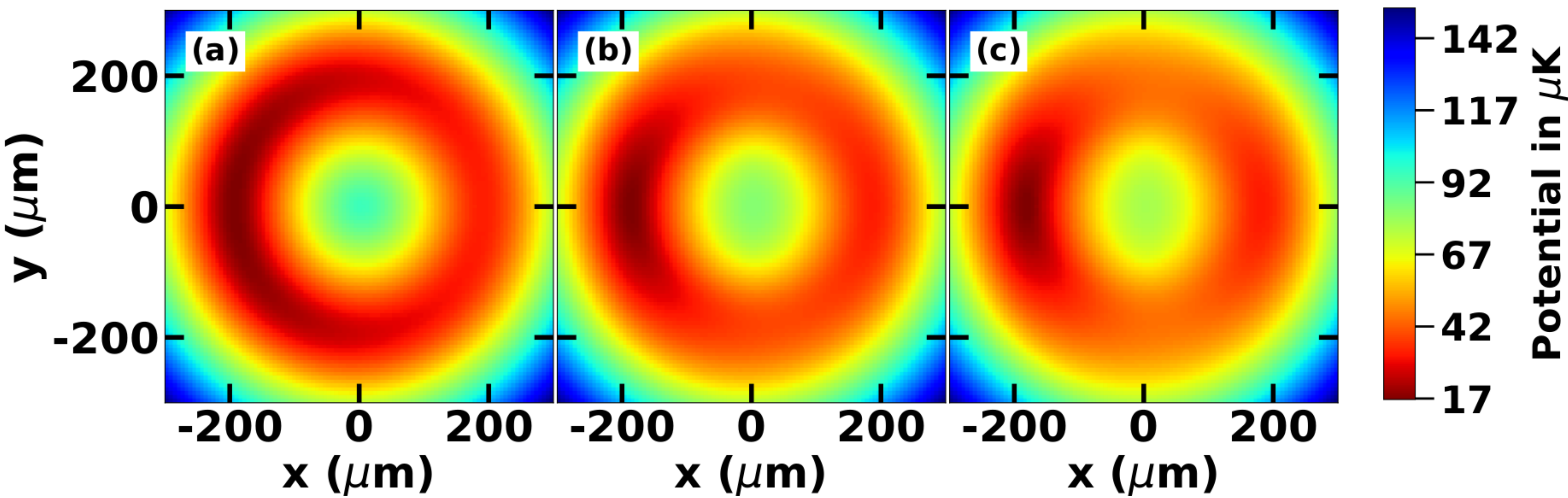}
 \caption{(Color online) The iso-potential contours of TAAP showing an arc type trapping potential and its further conversion into a $x$-directional asymmetric double well trap. The $x$-$y$ circularly polarized rf-field with $B_{rf}^{x} = B_{rf}^{y} = 700$ mG and a linearly $z$-polarized rf-field with $B_{rf}^{z} = 300$ mG are considered. The value of $B_{T}^{z}$ is taken as $1.3$ G for all plots (a), (b) and (c). Only $z$-TOP field is considered for plot (a), where as $y$-$z$ TOP fields are considered for plots (b) and (c). The values of $B_{T}^{y}$ are $B_{T}^{y} = 650$ mG for plot (b) and $B_{T}^{y} = 850$ mG for plot (c). The other parameters are $\omega_{rf} = 2\pi \times 1.5$ MHz, $\omega_{T} = 2\pi \times 7$ kHz and ${{B}_{q}^{\prime}} = 100$ G/cm.}
\label{Arc to xdw}
\end{figure}

Similarly, an arc type of atom trap can be  converted into an asymmetric $x$-directional double well when $z$-TOP field is changed to $y$-$z$ TOP field with a phase difference of $\frac{\pi}{2}$ between $y$- and $z$-components. These results are shown in Fig. \ref{Arc to xdw}. The $y$-direction TOP field component is increased and $z$-TOP field is kept at a fixed value. The separation between the wells is tunable by changing the rf field frequency $\omega_{rf}$ and quadrupole field gradient ${B}_{q}^{\prime}$. 

An arc type of trap is useful for unequal population around the circumference of a ring. Rotation of an arc can be used to study superfluidity \cite{Ramanathan2011,Gupta2005} and atom interferometry \cite{Navez_2016}. When arc type geometry is getting converted into a asymmetric double well, this conversion can be used to split the atom cloud in an unequal proportion which is a suitable geometry for study of tunnelling \cite{Albiez2005}.

\section{Conclusion} 
\label{sec:conclusion}

We have theoretically shown various kinds of atom trapping geometries by using TAAP scheme. It is shown that TAAP scheme has versatility to design and manipulate the potential for atom trapping with appropriate combination of static magnetic field, rf-fields and TOP fields. Various conversions among TAAP geometries, such as conversions between different directional double well traps, from ring trap to double well trap and from an arc trap to double well trap, have been discussed. These conversions between different trapping geometries can be useful to study the quantum degeneracy in low dimensions, dynamics of atom traps, matter-wave interferometry, superfluidity and tunnelling.
\section{Acknowledgement}
\label{sec:Acknowledgement}
We acknowledge Kavish Bhardwaj for useful discussion and a careful reading of the manuscript. Sourabh Sarkar acknowledges the financial support by Raja Ramanna Centre for Advanced Technology, Indore under Homi Bhabha National Institute (HBNI) programme.


\begin{thebibliography}{34}
\expandafter\ifx\csname natexlab\endcsname\relax\def\natexlab#1{#1}\fi
\expandafter\ifx\csname bibnamefont\endcsname\relax
  \def\bibnamefont#1{#1}\fi
\expandafter\ifx\csname bibfnamefont\endcsname\relax
  \def\bibfnamefont#1{#1}\fi
\expandafter\ifx\csname citenamefont\endcsname\relax
  \def\citenamefont#1{#1}\fi
\expandafter\ifx\csname url\endcsname\relax
  \def\url#1{\texttt{#1}}\fi
\expandafter\ifx\csname urlprefix\endcsname\relax\def\urlprefix{URL }\fi
\providecommand{\bibinfo}[2]{#2}
\providecommand{\eprint}[2][]{\url{#2}}

\bibitem[{\citenamefont{Metcalf}(1999)}]{Metcalf1999}
\bibinfo{author}{\bibfnamefont{H.~J.} \bibnamefont{Metcalf}},
  \emph{\bibinfo{title}{Laser Cooling and Trapping}}
  (\bibinfo{publisher}{Springer}, \bibinfo{year}{1999}).

\bibitem[{\citenamefont{Albiez et~al.}(2005)\citenamefont{Albiez, Gati,
  F\"olling, Hunsmann, Cristiani, and Oberthaler}}]{Albiez2005}
\bibinfo{author}{\bibfnamefont{M.}~\bibnamefont{Albiez}},
  \bibinfo{author}{\bibfnamefont{R.}~\bibnamefont{Gati}},
  \bibinfo{author}{\bibfnamefont{J.}~\bibnamefont{F\"olling}},
  \bibinfo{author}{\bibfnamefont{S.}~\bibnamefont{Hunsmann}},
  \bibinfo{author}{\bibfnamefont{M.}~\bibnamefont{Cristiani}},
  \bibnamefont{and} \bibinfo{author}{\bibfnamefont{M.~K.}
  \bibnamefont{Oberthaler}}, \bibinfo{journal}{\emph{Phys. Rev. Lett.}}
  \textbf{\bibinfo{volume}{95}}, \bibinfo{pages}{010402}
  (\bibinfo{year}{2005}),
  \urlprefix\url{https://link.aps.org/doi/10.1103/PhysRevLett.95.010402}.

\bibitem[{\citenamefont{Ramanathan et~al.}(2011)\citenamefont{Ramanathan,
  Wright, Muniz, Zelan, Hill, Lobb, Helmerson, Phillips, and
  Campbell}}]{Ramanathan2011}
\bibinfo{author}{\bibfnamefont{A.}~\bibnamefont{Ramanathan}},
  \bibinfo{author}{\bibfnamefont{K.~C.} \bibnamefont{Wright}},
  \bibinfo{author}{\bibfnamefont{S.~R.} \bibnamefont{Muniz}},
  \bibinfo{author}{\bibfnamefont{M.}~\bibnamefont{Zelan}},
  \bibinfo{author}{\bibfnamefont{W.~T.} \bibnamefont{Hill}},
  \bibinfo{author}{\bibfnamefont{C.~J.} \bibnamefont{Lobb}},
  \bibinfo{author}{\bibfnamefont{K.}~\bibnamefont{Helmerson}},
  \bibinfo{author}{\bibfnamefont{W.~D.} \bibnamefont{Phillips}},
  \bibnamefont{and} \bibinfo{author}{\bibfnamefont{G.~K.}
  \bibnamefont{Campbell}}, \bibinfo{journal}{\emph{Phys. Rev. Lett.}}
  \textbf{\bibinfo{volume}{106}}, \bibinfo{pages}{130401}
  (\bibinfo{year}{2011}),
  \urlprefix\url{https://link.aps.org/doi/10.1103/PhysRevLett.106.130401}.

\bibitem[{\citenamefont{Schumm et~al.}(2005)\citenamefont{Schumm, Hofferberth,
  Andersson, Wildermuth, Groth, Bar-Joseph, Schmiedmayer, and
  Kr{\"u}ger}}]{Schumm2005}
\bibinfo{author}{\bibfnamefont{T.}~\bibnamefont{Schumm}},
  \bibinfo{author}{\bibfnamefont{S.}~\bibnamefont{Hofferberth}},
  \bibinfo{author}{\bibfnamefont{L.~M.} \bibnamefont{Andersson}},
  \bibinfo{author}{\bibfnamefont{S.}~\bibnamefont{Wildermuth}},
  \bibinfo{author}{\bibfnamefont{S.}~\bibnamefont{Groth}},
  \bibinfo{author}{\bibfnamefont{I.}~\bibnamefont{Bar-Joseph}},
  \bibinfo{author}{\bibfnamefont{J.}~\bibnamefont{Schmiedmayer}},
  \bibnamefont{and}
  \bibinfo{author}{\bibfnamefont{P.}~\bibnamefont{Kr{\"u}ger}},
  \bibinfo{journal}{\emph{Nature Physics}} \textbf{\bibinfo{volume}{1}},
  \bibinfo{pages}{57} (\bibinfo{year}{2005}), ISSN \bibinfo{issn}{1745-2481},
  \urlprefix\url{https://doi.org/10.1038/nphys125}.

\bibitem[{\citenamefont{Bertoldi et~al.}(2006)\citenamefont{Bertoldi,
  Lamporesi, Cacciapuoti, de~Angelis, Fattori, Petelski, Peters, Prevedelli,
  Stuhler, and Tino}}]{BertoldiA.2006}
\bibinfo{author}{\bibfnamefont{A.}~\bibnamefont{Bertoldi}},
  \bibinfo{author}{\bibfnamefont{G.}~\bibnamefont{Lamporesi}},
  \bibinfo{author}{\bibfnamefont{L.}~\bibnamefont{Cacciapuoti}},
  \bibinfo{author}{\bibfnamefont{M.}~\bibnamefont{de~Angelis}},
  \bibinfo{author}{\bibfnamefont{M.}~\bibnamefont{Fattori}},
  \bibinfo{author}{\bibfnamefont{T.}~\bibnamefont{Petelski}},
  \bibinfo{author}{\bibfnamefont{A.}~\bibnamefont{Peters}},
  \bibinfo{author}{\bibfnamefont{M.}~\bibnamefont{Prevedelli}},
  \bibinfo{author}{\bibfnamefont{J.}~\bibnamefont{Stuhler}}, \bibnamefont{and}
  \bibinfo{author}{\bibfnamefont{G.~M.} \bibnamefont{Tino}},
  \bibinfo{journal}{\emph{Eur. Phys. J. D}} \textbf{\bibinfo{volume}{40}},
  \bibinfo{pages}{271} (\bibinfo{year}{2006}),
  \urlprefix\url{https://doi.org/10.1140/epjd/e2006-00212-2}.

\bibitem[{\citenamefont{Peters et~al.}(2001)\citenamefont{Peters, Chung, and
  Chu}}]{Peters2001}
\bibinfo{author}{\bibfnamefont{A.}~\bibnamefont{Peters}},
  \bibinfo{author}{\bibfnamefont{K.~Y.} \bibnamefont{Chung}}, \bibnamefont{and}
  \bibinfo{author}{\bibfnamefont{S.}~\bibnamefont{Chu}},
  \bibinfo{journal}{\emph{Metrologia}} \textbf{\bibinfo{volume}{38}},
  \bibinfo{pages}{25} (\bibinfo{year}{2001}),
  \urlprefix\url{https://doi.org/10.1088%2F0026-1394%2F38%2F1%2F4}.

\bibitem[{\citenamefont{M\"uller et~al.}(2009)\citenamefont{M\"uller, Gilowski,
  Zaiser, Berg, Schubert, Wendrich, Ertmer, and Rasel}}]{MuellerT.2009}
\bibinfo{author}{\bibfnamefont{T.}~\bibnamefont{M\"uller}},
  \bibinfo{author}{\bibfnamefont{M.}~\bibnamefont{Gilowski}},
  \bibinfo{author}{\bibfnamefont{M.}~\bibnamefont{Zaiser}},
  \bibinfo{author}{\bibfnamefont{P.}~\bibnamefont{Berg}},
  \bibinfo{author}{\bibfnamefont{C.}~\bibnamefont{Schubert}},
  \bibinfo{author}{\bibfnamefont{T.}~\bibnamefont{Wendrich}},
  \bibinfo{author}{\bibfnamefont{W.}~\bibnamefont{Ertmer}}, \bibnamefont{and}
  \bibinfo{author}{\bibfnamefont{E.~M.} \bibnamefont{Rasel}},
  \bibinfo{journal}{\emph{Eur. Phys. J. D}} \textbf{\bibinfo{volume}{53}},
  \bibinfo{pages}{273} (\bibinfo{year}{2009}),
  \urlprefix\url{https://doi.org/10.1140/epjd/e2009-00139-0}.

\bibitem[{\citenamefont{Mark et~al.}(2011)\citenamefont{Mark, Haller, Lauber,
  Danzl, Daley, and N\"agerl}}]{Mark2011}
\bibinfo{author}{\bibfnamefont{M.~J.} \bibnamefont{Mark}},
  \bibinfo{author}{\bibfnamefont{E.}~\bibnamefont{Haller}},
  \bibinfo{author}{\bibfnamefont{K.}~\bibnamefont{Lauber}},
  \bibinfo{author}{\bibfnamefont{J.~G.} \bibnamefont{Danzl}},
  \bibinfo{author}{\bibfnamefont{A.~J.} \bibnamefont{Daley}}, \bibnamefont{and}
  \bibinfo{author}{\bibfnamefont{H.-C.} \bibnamefont{N\"agerl}},
  \bibinfo{journal}{\emph{Phys. Rev. Lett.}} \textbf{\bibinfo{volume}{107}},
  \bibinfo{pages}{175301} (\bibinfo{year}{2011}),
  \urlprefix\url{https://link.aps.org/doi/10.1103/PhysRevLett.107.175301}.

\bibitem[{\citenamefont{Ryu et~al.}(2013)\citenamefont{Ryu, Blackburn, Blinova,
  and Boshier}}]{CRYU2013}
\bibinfo{author}{\bibfnamefont{C.}~\bibnamefont{Ryu}},
  \bibinfo{author}{\bibfnamefont{P.~W.} \bibnamefont{Blackburn}},
  \bibinfo{author}{\bibfnamefont{A.~A.} \bibnamefont{Blinova}},
  \bibnamefont{and} \bibinfo{author}{\bibfnamefont{M.~G.}
  \bibnamefont{Boshier}}, \bibinfo{journal}{\emph{Phys. Rev. Lett.}}
  \textbf{\bibinfo{volume}{111}}, \bibinfo{pages}{205301}
  (\bibinfo{year}{2013}),
  \urlprefix\url{https://link.aps.org/doi/10.1103/PhysRevLett.111.205301}.

\bibitem[{\citenamefont{Alzar}(2019)}]{Alzar2019}
\bibinfo{author}{\bibfnamefont{C.~L.~G.} \bibnamefont{Alzar}},
  \bibinfo{journal}{\emph{AVS Quantum Sci.}} \textbf{\bibinfo{volume}{1}},
  \bibinfo{pages}{014702} (\bibinfo{year}{2019}),
  \urlprefix\url{https://doi.org/10.1116/1.5142003}.

\bibitem[{\citenamefont{Wohlleben et~al.}(2001)\citenamefont{Wohlleben, Chevy,
  Madison, and Dalibard}}]{WohllebenW.2001}
\bibinfo{author}{\bibfnamefont{W.}~\bibnamefont{Wohlleben}},
  \bibinfo{author}{\bibfnamefont{F.}~\bibnamefont{Chevy}},
  \bibinfo{author}{\bibfnamefont{K.}~\bibnamefont{Madison}}, \bibnamefont{and}
  \bibinfo{author}{\bibfnamefont{J.}~\bibnamefont{Dalibard}},
  \bibinfo{journal}{\emph{Eur. Phys. J. D}} \textbf{\bibinfo{volume}{15}},
  \bibinfo{pages}{237} (\bibinfo{year}{2001}),
  \urlprefix\url{https://doi.org/10.1007/s100530170171}.

\bibitem[{\citenamefont{Merloti et~al.}(2013)\citenamefont{Merloti, Dubessy,
  Longchambon, Perrin, Pottie, Lorent, and Perrin}}]{Merloti2013}
\bibinfo{author}{\bibfnamefont{K.}~\bibnamefont{Merloti}},
  \bibinfo{author}{\bibfnamefont{R.}~\bibnamefont{Dubessy}},
  \bibinfo{author}{\bibfnamefont{L.}~\bibnamefont{Longchambon}},
  \bibinfo{author}{\bibfnamefont{A.}~\bibnamefont{Perrin}},
  \bibinfo{author}{\bibfnamefont{P.-E.} \bibnamefont{Pottie}},
  \bibinfo{author}{\bibfnamefont{V.}~\bibnamefont{Lorent}}, \bibnamefont{and}
  \bibinfo{author}{\bibfnamefont{H.}~\bibnamefont{Perrin}},
  \bibinfo{journal}{\emph{New Journal of Physics}}
  \textbf{\bibinfo{volume}{15}}, \bibinfo{pages}{033007}
  (\bibinfo{year}{2013}),
  \urlprefix\url{https://doi.org/10.1088/1367-2630/15/3/033007}.

\bibitem[{\citenamefont{Chakraborty et~al.}(2016)\citenamefont{Chakraborty,
  Mishra, Ram, Tiwari, and Rawat}}]{Chakraborty2016}
\bibinfo{author}{\bibfnamefont{A.}~\bibnamefont{Chakraborty}},
  \bibinfo{author}{\bibfnamefont{S.~R.} \bibnamefont{Mishra}},
  \bibinfo{author}{\bibfnamefont{S.~P.} \bibnamefont{Ram}},
  \bibinfo{author}{\bibfnamefont{S.~K.} \bibnamefont{Tiwari}},
  \bibnamefont{and} \bibinfo{author}{\bibfnamefont{H.~S.} \bibnamefont{Rawat}},
  \bibinfo{journal}{\emph{Journal of Physics B: Atomic, Molecular and Optical
  Physics}} \textbf{\bibinfo{volume}{49}}, \bibinfo{pages}{075304}
  (\bibinfo{year}{2016}),
  \urlprefix\url{https://doi.org/10.1088%2F0953-4075%2F49%2F7%2F075304}.

\bibitem[{\citenamefont{Grimm et~al.}(2000)\citenamefont{Grimm, Weidemüller,
  and Ovchinnikov}}]{GRIMM200095}
\bibinfo{author}{\bibfnamefont{R.}~\bibnamefont{Grimm}},
  \bibinfo{author}{\bibfnamefont{M.}~\bibnamefont{Weidemüller}},
  \bibnamefont{and} \bibinfo{author}{\bibfnamefont{Y.~B.}
  \bibnamefont{Ovchinnikov}} (\bibinfo{publisher}{Academic Press},
  \bibinfo{year}{2000}), vol.~\bibinfo{volume}{42} of
  \emph{\bibinfo{series}{Advances In Atomic, Molecular, and Optical Physics}},
  pp. \bibinfo{pages}{95 -- 170},
  \urlprefix\url{http://www.sciencedirect.com/science/article/pii/S1049250X0860186X}.

\bibitem[{\citenamefont{Zobay and Garraway}(2001)}]{Zobay2001}
\bibinfo{author}{\bibfnamefont{O.}~\bibnamefont{Zobay}} \bibnamefont{and}
  \bibinfo{author}{\bibfnamefont{B.~M.} \bibnamefont{Garraway}},
  \bibinfo{journal}{\emph{Phys. Rev. Lett.}} \textbf{\bibinfo{volume}{86}},
  \bibinfo{pages}{1195} (\bibinfo{year}{2001}),
  \urlprefix\url{https://link.aps.org/doi/10.1103/PhysRevLett.86.1195}.

\bibitem[{\citenamefont{Chakraborty and Mishra}(2014)}]{Chakraborty2014}
\bibinfo{author}{\bibfnamefont{A.}~\bibnamefont{Chakraborty}} \bibnamefont{and}
  \bibinfo{author}{\bibfnamefont{S.~R.} \bibnamefont{Mishra}},
  \bibinfo{journal}{\emph{Journal of the Korean Physical Society}}
  \textbf{\bibinfo{volume}{65}}, \bibinfo{pages}{1324} (\bibinfo{year}{2014}),
  \urlprefix\url{https://doi.org/10.3938/jkps.65.1324}.

\bibitem[{\citenamefont{Heathcote et~al.}(2008)\citenamefont{Heathcote, Nugent,
  Sheard, and Foot}}]{Heathcote2008}
\bibinfo{author}{\bibfnamefont{W.~H.} \bibnamefont{Heathcote}},
  \bibinfo{author}{\bibfnamefont{E.}~\bibnamefont{Nugent}},
  \bibinfo{author}{\bibfnamefont{B.~T.} \bibnamefont{Sheard}},
  \bibnamefont{and} \bibinfo{author}{\bibfnamefont{C.~J.} \bibnamefont{Foot}},
  \bibinfo{journal}{\emph{New Journal of Physics}}
  \textbf{\bibinfo{volume}{10}}, \bibinfo{pages}{043012}
  (\bibinfo{year}{2008}),
  \urlprefix\url{https://doi.org/10.1088%2F1367-2630%2F10%2F4%2F043012}.

\bibitem[{\citenamefont{Hofferberth et~al.}(2006)\citenamefont{Hofferberth,
  Lesanovsky, Fischer, Verdu, and Schmiedmayer}}]{Hofferberth2006}
\bibinfo{author}{\bibfnamefont{S.}~\bibnamefont{Hofferberth}},
  \bibinfo{author}{\bibfnamefont{I.}~\bibnamefont{Lesanovsky}},
  \bibinfo{author}{\bibfnamefont{B.}~\bibnamefont{Fischer}},
  \bibinfo{author}{\bibfnamefont{J.}~\bibnamefont{Verdu}}, \bibnamefont{and}
  \bibinfo{author}{\bibfnamefont{J.}~\bibnamefont{Schmiedmayer}},
  \bibinfo{journal}{\emph{Nature Physics}} \textbf{\bibinfo{volume}{2}},
  \bibinfo{pages}{710} (\bibinfo{year}{2006}),
  \urlprefix\url{https://doi.org/10.1038/nphys420}.

\bibitem[{\citenamefont{Sherlock et~al.}(2011)\citenamefont{Sherlock,
  Gildemeister, Owen, Nugent, and Foot}}]{Sherlock2011}
\bibinfo{author}{\bibfnamefont{B.~E.} \bibnamefont{Sherlock}},
  \bibinfo{author}{\bibfnamefont{M.}~\bibnamefont{Gildemeister}},
  \bibinfo{author}{\bibfnamefont{E.}~\bibnamefont{Owen}},
  \bibinfo{author}{\bibfnamefont{E.}~\bibnamefont{Nugent}}, \bibnamefont{and}
  \bibinfo{author}{\bibfnamefont{C.~J.} \bibnamefont{Foot}},
  \bibinfo{journal}{\emph{Phys. Rev. A}} \textbf{\bibinfo{volume}{83}},
  \bibinfo{pages}{043408} (\bibinfo{year}{2011}),
  \urlprefix\url{https://link.aps.org/doi/10.1103/PhysRevA.83.043408}.

\bibitem[{\citenamefont{Easwaran et~al.}(2010)\citenamefont{Easwaran,
  Longchambon, Pottie, Lorent, Perrin, and Garraway}}]{Easwaran2010}
\bibinfo{author}{\bibfnamefont{R.~K.} \bibnamefont{Easwaran}},
  \bibinfo{author}{\bibfnamefont{L.}~\bibnamefont{Longchambon}},
  \bibinfo{author}{\bibfnamefont{P.-E.} \bibnamefont{Pottie}},
  \bibinfo{author}{\bibfnamefont{V.}~\bibnamefont{Lorent}},
  \bibinfo{author}{\bibfnamefont{H.}~\bibnamefont{Perrin}}, \bibnamefont{and}
  \bibinfo{author}{\bibfnamefont{B.~M.} \bibnamefont{Garraway}},
  \bibinfo{journal}{\emph{Journal of Physics B: Atomic, Molecular and Optical
  Physics}} \textbf{\bibinfo{volume}{43}}, \bibinfo{pages}{065302}
  (\bibinfo{year}{2010}),
  \urlprefix\url{https://doi.org/10.1088%2F0953-4075%2F43%2F6%2F065302}.

\bibitem[{\citenamefont{Morizot et~al.}(2006)\citenamefont{Morizot, Colombe,
  Lorent, Perrin, and Garraway}}]{Morizot2006}
\bibinfo{author}{\bibfnamefont{O.}~\bibnamefont{Morizot}},
  \bibinfo{author}{\bibfnamefont{Y.}~\bibnamefont{Colombe}},
  \bibinfo{author}{\bibfnamefont{V.}~\bibnamefont{Lorent}},
  \bibinfo{author}{\bibfnamefont{H.}~\bibnamefont{Perrin}}, \bibnamefont{and}
  \bibinfo{author}{\bibfnamefont{B.~M.} \bibnamefont{Garraway}},
  \bibinfo{journal}{\emph{Phys. Rev. A}} \textbf{\bibinfo{volume}{74}},
  \bibinfo{pages}{023617} (\bibinfo{year}{2006}),
  \urlprefix\url{https://link.aps.org/doi/10.1103/PhysRevA.74.023617}.

\bibitem[{\citenamefont{Lesanovsky and von Klitzing}(2007)}]{Lesanovsky2007}
\bibinfo{author}{\bibfnamefont{I.}~\bibnamefont{Lesanovsky}} \bibnamefont{and}
  \bibinfo{author}{\bibfnamefont{W.}~\bibnamefont{von Klitzing}},
  \bibinfo{journal}{\emph{Phys. Rev. Lett.}} \textbf{\bibinfo{volume}{99}},
  \bibinfo{pages}{083001} (\bibinfo{year}{2007}),
  \urlprefix\url{https://link.aps.org/doi/10.1103/PhysRevLett.99.083001}.

\bibitem[{\citenamefont{Petrich et~al.}(1995)\citenamefont{Petrich, Anderson,
  Ensher, and Cornell}}]{Petrich1995}
\bibinfo{author}{\bibfnamefont{W.}~\bibnamefont{Petrich}},
  \bibinfo{author}{\bibfnamefont{M.~H.} \bibnamefont{Anderson}},
  \bibinfo{author}{\bibfnamefont{J.~R.} \bibnamefont{Ensher}},
  \bibnamefont{and} \bibinfo{author}{\bibfnamefont{E.~A.}
  \bibnamefont{Cornell}}, \bibinfo{journal}{\emph{Phys. Rev. Lett.}}
  \textbf{\bibinfo{volume}{74}}, \bibinfo{pages}{3352} (\bibinfo{year}{1995}),
  \urlprefix\url{https://link.aps.org/doi/10.1103/PhysRevLett.74.3352}.

\bibitem[{\citenamefont{Gildemeister et~al.}(2012)\citenamefont{Gildemeister,
  Sherlock, and Foot}}]{Gildemeister2012}
\bibinfo{author}{\bibfnamefont{M.}~\bibnamefont{Gildemeister}},
  \bibinfo{author}{\bibfnamefont{B.~E.} \bibnamefont{Sherlock}},
  \bibnamefont{and} \bibinfo{author}{\bibfnamefont{C.~J.} \bibnamefont{Foot}},
  \bibinfo{journal}{\emph{Phys. Rev. A}} \textbf{\bibinfo{volume}{85}},
  \bibinfo{pages}{053401} (\bibinfo{year}{2012}),
  \urlprefix\url{https://link.aps.org/doi/10.1103/PhysRevA.85.053401}.

\bibitem[{\citenamefont{Anderson et~al.}(1995)\citenamefont{Anderson, Ensher,
  Matthews, Wieman, and Cornell}}]{Anderson1995}
\bibinfo{author}{\bibfnamefont{M.~H.} \bibnamefont{Anderson}},
  \bibinfo{author}{\bibfnamefont{J.~R.} \bibnamefont{Ensher}},
  \bibinfo{author}{\bibfnamefont{M.~R.} \bibnamefont{Matthews}},
  \bibinfo{author}{\bibfnamefont{C.~E.} \bibnamefont{Wieman}},
  \bibnamefont{and} \bibinfo{author}{\bibfnamefont{E.~A.}
  \bibnamefont{Cornell}}, \bibinfo{journal}{\emph{Science}}
  \textbf{\bibinfo{volume}{269}}, \bibinfo{pages}{198} (\bibinfo{year}{1995}),
  ISSN \bibinfo{issn}{0036-8075},
  \eprint{https://science.sciencemag.org/content/269/5221/198.full.pdf},
  \urlprefix\url{https://science.sciencemag.org/content/269/5221/198}.

\bibitem[{\citenamefont{Davis et~al.}(1995)\citenamefont{Davis, Mewes, Andrews,
  van Druten, Durfee, Kurn, and Ketterle}}]{Davis1995}
\bibinfo{author}{\bibfnamefont{K.~B.} \bibnamefont{Davis}},
  \bibinfo{author}{\bibfnamefont{M.~O.} \bibnamefont{Mewes}},
  \bibinfo{author}{\bibfnamefont{M.~R.} \bibnamefont{Andrews}},
  \bibinfo{author}{\bibfnamefont{N.~J.} \bibnamefont{van Druten}},
  \bibinfo{author}{\bibfnamefont{D.~S.} \bibnamefont{Durfee}},
  \bibinfo{author}{\bibfnamefont{D.~M.} \bibnamefont{Kurn}}, \bibnamefont{and}
  \bibinfo{author}{\bibfnamefont{W.}~\bibnamefont{Ketterle}},
  \bibinfo{journal}{\emph{Phys. Rev. Lett.}} \textbf{\bibinfo{volume}{75}},
  \bibinfo{pages}{3969} (\bibinfo{year}{1995}),
  \urlprefix\url{https://link.aps.org/doi/10.1103/PhysRevLett.75.3969}.

\bibitem[{\citenamefont{Hofferberth et~al.}(2007)\citenamefont{Hofferberth,
  Fischer, Schumm, Schmiedmayer, and Lesanovsky}}]{Hofferberth2007}
\bibinfo{author}{\bibfnamefont{S.}~\bibnamefont{Hofferberth}},
  \bibinfo{author}{\bibfnamefont{B.}~\bibnamefont{Fischer}},
  \bibinfo{author}{\bibfnamefont{T.}~\bibnamefont{Schumm}},
  \bibinfo{author}{\bibfnamefont{J.}~\bibnamefont{Schmiedmayer}},
  \bibnamefont{and}
  \bibinfo{author}{\bibfnamefont{I.}~\bibnamefont{Lesanovsky}},
  \bibinfo{journal}{\emph{Phys. Rev. A}} \textbf{\bibinfo{volume}{76}},
  \bibinfo{pages}{013401} (\bibinfo{year}{2007}),
  \urlprefix\url{https://link.aps.org/doi/10.1103/PhysRevA.76.013401}.

\bibitem[{\citenamefont{Kasevich and Chu}(1991)}]{Kasevich1991}
\bibinfo{author}{\bibfnamefont{M.}~\bibnamefont{Kasevich}} \bibnamefont{and}
  \bibinfo{author}{\bibfnamefont{S.}~\bibnamefont{Chu}},
  \bibinfo{journal}{\emph{Phys. Rev. Lett.}} \textbf{\bibinfo{volume}{67}},
  \bibinfo{pages}{181} (\bibinfo{year}{1991}),
  \urlprefix\url{https://link.aps.org/doi/10.1103/PhysRevLett.67.181}.

\bibitem[{\citenamefont{Canuel et~al.}(2006)\citenamefont{Canuel, Leduc,
  Holleville, Gauguet, Fils, Virdis, Clairon, Dimarcq, Bord\'e, Landragin
  et~al.}}]{Canuel2006}
\bibinfo{author}{\bibfnamefont{B.}~\bibnamefont{Canuel}},
  \bibinfo{author}{\bibfnamefont{F.}~\bibnamefont{Leduc}},
  \bibinfo{author}{\bibfnamefont{D.}~\bibnamefont{Holleville}},
  \bibinfo{author}{\bibfnamefont{A.}~\bibnamefont{Gauguet}},
  \bibinfo{author}{\bibfnamefont{J.}~\bibnamefont{Fils}},
  \bibinfo{author}{\bibfnamefont{A.}~\bibnamefont{Virdis}},
  \bibinfo{author}{\bibfnamefont{A.}~\bibnamefont{Clairon}},
  \bibinfo{author}{\bibfnamefont{N.}~\bibnamefont{Dimarcq}},
  \bibinfo{author}{\bibfnamefont{C.~J.} \bibnamefont{Bord\'e}},
  \bibinfo{author}{\bibfnamefont{A.}~\bibnamefont{Landragin}},
  \bibnamefont{et~al.}, \bibinfo{journal}{\emph{Phys. Rev. Lett.}}
  \textbf{\bibinfo{volume}{97}}, \bibinfo{pages}{010402}
  (\bibinfo{year}{2006}),
  \urlprefix\url{https://link.aps.org/doi/10.1103/PhysRevLett.97.010402}.

\bibitem[{\citenamefont{Carnal and Mlynek}(1991)}]{Carnal1991}
\bibinfo{author}{\bibfnamefont{O.}~\bibnamefont{Carnal}} \bibnamefont{and}
  \bibinfo{author}{\bibfnamefont{J.}~\bibnamefont{Mlynek}},
  \bibinfo{journal}{\emph{Phys. Rev. Lett.}} \textbf{\bibinfo{volume}{66}},
  \bibinfo{pages}{2689} (\bibinfo{year}{1991}),
  \urlprefix\url{https://link.aps.org/doi/10.1103/PhysRevLett.66.2689}.

\bibitem[{\citenamefont{Gildemeister}(2010)}]{Gildemeister2010a}
\bibinfo{author}{\bibfnamefont{M.}~\bibnamefont{Gildemeister}}, Ph.D. thesis,
  \bibinfo{school}{University of Oxford} (\bibinfo{year}{2010}).

\bibitem[{\citenamefont{Gildemeister et~al.}(2010)\citenamefont{Gildemeister,
  Nugent, Sherlock, Kubasik, Sheard, and Foot}}]{Gildemeister2010}
\bibinfo{author}{\bibfnamefont{M.}~\bibnamefont{Gildemeister}},
  \bibinfo{author}{\bibfnamefont{E.}~\bibnamefont{Nugent}},
  \bibinfo{author}{\bibfnamefont{B.~E.} \bibnamefont{Sherlock}},
  \bibinfo{author}{\bibfnamefont{M.}~\bibnamefont{Kubasik}},
  \bibinfo{author}{\bibfnamefont{B.~T.} \bibnamefont{Sheard}},
  \bibnamefont{and} \bibinfo{author}{\bibfnamefont{C.~J.} \bibnamefont{Foot}},
  \bibinfo{journal}{\emph{Phys. Rev. A}} \textbf{\bibinfo{volume}{81}},
  \bibinfo{pages}{031402} (\bibinfo{year}{2010}),
  \urlprefix\url{https://link.aps.org/doi/10.1103/PhysRevA.81.031402}.

\bibitem[{\citenamefont{Navez et~al.}(2016)\citenamefont{Navez, Pandey, Mas,
  Poulios, Fernholz, and von Klitzing}}]{Navez_2016}
\bibinfo{author}{\bibfnamefont{P.}~\bibnamefont{Navez}},
  \bibinfo{author}{\bibfnamefont{S.}~\bibnamefont{Pandey}},
  \bibinfo{author}{\bibfnamefont{H.}~\bibnamefont{Mas}},
  \bibinfo{author}{\bibfnamefont{K.}~\bibnamefont{Poulios}},
  \bibinfo{author}{\bibfnamefont{T.}~\bibnamefont{Fernholz}}, \bibnamefont{and}
  \bibinfo{author}{\bibfnamefont{W.}~\bibnamefont{von Klitzing}},
  \bibinfo{journal}{\emph{New Journal of Physics}}
  \textbf{\bibinfo{volume}{18}}, \bibinfo{pages}{075014}
  (\bibinfo{year}{2016}),
  \urlprefix\url{https://doi.org/10.1088/1367-2630/18/7/075014}.

\bibitem[{\citenamefont{Gupta et~al.}(2005)\citenamefont{Gupta, Murch, Moore,
  Purdy, and Stamper-Kurn}}]{Gupta2005}
\bibinfo{author}{\bibfnamefont{S.}~\bibnamefont{Gupta}},
  \bibinfo{author}{\bibfnamefont{K.~W.} \bibnamefont{Murch}},
  \bibinfo{author}{\bibfnamefont{K.~L.} \bibnamefont{Moore}},
  \bibinfo{author}{\bibfnamefont{T.~P.} \bibnamefont{Purdy}}, \bibnamefont{and}
  \bibinfo{author}{\bibfnamefont{D.~M.} \bibnamefont{Stamper-Kurn}},
  \bibinfo{journal}{\emph{Phys. Rev. Lett.}} \textbf{\bibinfo{volume}{95}},
  \bibinfo{pages}{143201} (\bibinfo{year}{2005}),
  \urlprefix\url{https://link.aps.org/doi/10.1103/PhysRevLett.95.143201}.

\end{thebibliography}
\end{document}